\documentclass[useAMS,usenatbib]{mn2e}

\usepackage{amsmath}
\usepackage{amssymb}

\usepackage{epsfig}
\usepackage{graphicx}
\usepackage{subfigure}
\usepackage{longtable}
\usepackage{multirow}
\usepackage{xcolor}

\newcommand{\iratio}{$I_{\rm 7.0}/I_{\rm 6.7}$}

\newcommand{\suzaku}{\textit{Suzaku}}

\newcommand{\swift}{\textit{Swift}}
\newcommand{\xmm}{\textit{XMM-Newton}}
\newcommand{\tmax}{$T_{\rm max}$}
\newcommand{\mwd}{$M_{\rm WD}$}

\title{V1082-Sgr: A magnetic cataclysmic variable with a lobe-filling companion star}

\author[Xu, Shao \& Li]{Xiaojie Xu$^{1,2}$\thanks{E-mail:
xuxj@nju.edu.cn}, Yong Shao$^{1,2}$ and Xiang-Dong Li$^{1,2}$\\
$^{1}$School of Astronomy and Space Science, Nanjing University, Nanjing 210046, China\\
$^{2}$Key Laboratory of Modern Astronomy and Astrophysics (Nanjing University), Ministry of
Education, Nanjing 210046, China}

\begin{document}

\date{Accepted 1988 December 15. Received 1988 December 14; in original form 1988 October 11}

\pagerange{\pageref{firstpage}--\pageref{lastpage}} \pubyear{2002}

\maketitle

\label{firstpage}

\begin{abstract}

V1082 Sgr is a cataclysmic variable with accretion luminosity above $10^{34}$erg s$^{-1}$, indicating a mass transfer rate above $10^{-9}M_{\odot}$yr$^{-1}$. However, its K type companion was suggested to be under-filling its Roche lobe (RL), making the high mass transfer rate a mystery. In this work we propose a possible model to explain this discrepancy. The system is proposed to be an intermediate polar, with its K type companion filling its RL. The mass of the white dwarf star is evaluated to be $0.77\pm0.11M_{\odot}$ from both X-ray continuum fitting and Fe line flux ratio measurements. We make numerical simulations to search for the possible progenitors of the system. The results show that a binary  with an initial 1.5--2.5$M_{\odot}$ companion in a 1--2 day orbit (or an initial 1.0--1.4$M_{\odot}$ companion in a 3.2--4.1 day orbit) may naturally evolve to a cataclysmic variable with a $\sim 0.55 \pm 0.11M_{\odot}$, Roche-lobe filling companion in a 0.86 day orbit. The effective temperature of the donor star, the mass transfer rate, and the derived V band magnitude are all consistent with previous observations. 
\end{abstract}


\section{Introduction}
Cataclysmic variables (CVs) are close binaries, consisting of a white dwarf (WD) accreting matter from its main sequence (MS) or sub-giant companion via Roche-lobe overflow (RLOF) or capture of the stellar wind (see \citealt{mukai2017} for a recent review). CVs can be classified into magnetic CVs (mCVs, including polars and intermediate polars (IPs)) and non-magnetic ones (normal CVs), according to the magnetic field strengths of the WDs. 
In normal CVs, accretion usually occurs via an accretion disk that extends to the surface of the WDs.
In mCVs, the magnetic fields of the WDs are so strong ($B\sim10^{6}$ G for IPs, and $\gtrsim10^{7}$ G for polars), that part or all of the accretion disk is disrupted, and the accreted matter is forced to move along the magnetic lines and settle in the magnetic poles of the WD. 
A standing shock is formed near the surface of the WD, and the accreted matter is heated. The subsequent cooling of the matter (including both bremsstrahlung and line transitions) gives rise to X-ray emission \citep{frank2002}. 

The X-ray spectra of polars are relatively soft, probably due to the dominance of cyclotron cooling in these objects.
The X-ray spectra of IPs, on the other hand, are relatively hard and can be well characterized by a multi-temperature optical thin thermal plasma model (mkcflow in xspec), sometimes with complex absorption which are believed to be intrinsic to this kind of CVs. 
Previous works have shown the typical 0.5-10 keV luminosities of IPs are in the range of $10^{32-34}$ erg s$^{-1}$ \citep[e.g.,][]{mukai2017}. The X-ray luminosity ($L_{\rm X}$) can be related to the white dwarf's accretion rate ($\dot{M}_{\rm acc}$) by $L_{\rm X}=\eta~G~M_{\rm WD}~\dot{M}_{\rm acc}/R_{\rm WD}$, where G is the gravitational constant, $M_{\rm WD}$ and $R_{\rm WD}$ are the mass and radius of the white dwarf, and $\eta$ is the percentage of energy released in X-ray form, which was set to be $0.09$ (or $0.7$) for the 2--8 keV (or the 0.1-100 keV) energy range \citep{ruiter2006,suleimanov2019}. Such high luminosities indicate a mass accretion rate in the order of  $10^{-9}M_{\odot}$yr$^{-1}$, which is one order of magnitude less than the companion mass loss rate predictions by binary evolution theories \citep[e.g.,][]{suleimanov2019}. 

V1082 Sgr (V1082 here after, also known as Swift J1907.3$−$2050) belongs to the family of mCVs. 
It was suggested to be a symbiotic star by \citet{cie1998}, and then classified as a mCV due to its hard X-ray spectra \citep{mukai2009}. Its orbital period was determined to be $20.82$ hr \citep{thor2010}. The WD mass was later derived to be $M_{\rm WD}=0.64\pm0.04M_{\odot}$ from the correlation between \mwd\ and the maximum temperature (\tmax), which was measured from the 3-50 keV X-ray spectra \citep{berna2013}. 
With a distance of $669\pm13$ pc, its peak 0.5-10 keV luminosity can reach $10^{34}$ erg s$^{-1}$ (implying a mass transfer rate of $\sim10^{-9}M_{\odot}$yr$^{-1}$, e.g., \citealt{berna2013}). The strong variability in both X-ray and optical bands may reflect the sudden changes in the accretion rates \citep{berna2013}. 
Recent optical observations revealed that the companion of V1082 is a K type star whose mass and radius were calculated to be $0.73\pm0.04M_{\odot}$ and $1.16\pm0.11R_{\odot}$, respectively \citep{tov2016,tov2018a,tov2018b}. Compared with the derived RL radius of $1.66\pm0.05R_{\odot}$ \citep{tov2018b}, it was suggested that the companion star was not filling its RL. 

How can a RL under-filling, K type MS star maintain a mass transfer rate of $10^{-9}M_{\odot}$yr$^{-1}$? 
It was suggested that V1082 could be a pre-polar, where the wind from the companion are directly channeled to the WD via its magnetic lines, a.k.a, the `bottle neck' mechanism \citep{parsons2013,tov2016,tov2018b}.
\citet{wood2002}, and \citet{ignace2010} suggested that the stellar wind loss rate of a fast spinning K type MS star may be enhanced and reach the order of $\sim10^{-11}M_{\odot}$yr$^{-1}$. However, this value is still two orders of magnitude lower than what is required to power the $10^{34}$ erg s$^{-1}$ luminosity.
What's more, the derived radius ($1.16\pm0.11R_{\odot}$) of the companion by \citet{tov2018b} is too large for a $0.73\pm0.04M_{\odot}$ MS star. \citet{tov2018b} proposed a scenario in which the companion has evolved off MS a little bit, and a thermal timescale mass transfer (TTMT) may have occurred. However, this mechanism is contradicted with the MS assumption, and could not explain why the companion is not filling the RL in present.

We notice that the interpretation of the companion mass and radius depends mostly on the WD mass, which was determined by the joint fitting of \xmm\ (3-10 keV) and \swift\ BAT (10-50 keV) data \citep{berna2013}. However, these observations were not taken simultaneously. Considering the fact that V1082 is a highly variable source \citep{berna2013}, the cross-calibration of X-ray data from different instruments in different time could be a problem and might have brought uncertainties for the inferred parameters, e.g., the mass of the WD. Thus, a re-evaluation of the WD mass is needed to better understand the nature of V1082.

In this work, we propose an mCV, more specifically, an IP model as a possible explanation for V1082. The IP model is preferred compared to the polar model, because the system shows a hard X-ray spectrum up to 40 keV, which is in contrast to polars which generally have soft X-ray spectra. In this model, the companion is a lobe-filling star, which has evolved off the MS. Our model is based on the re-evaluation of the WD mass of V1082 from archival \xmm\ and \suzaku\ observations. The evolution of the binary is then re-constructed with numerical simulation. We describe the X-ray data and analysis methods in Section 2, and present the derived system parameters in Section 3. We present the numerical simulation to explore the possible progenitors of V1082 in Section 4. Finally we summarize our findings with a brief discussion in Section 5. All uncertainties in this work are given in 90\% confidence level.

\section{Data Analysis}
V1082 has been observed by \suzaku, \xmm\ and \swift. We start our investigation with the \suzaku\ observation to ensure the simultaneous coverage of wide energy range (3-10 and 12-50 keV, with XIS and HXD instruments on board \suzaku, respectively) of the source spectra. The \xmm\ pn and MOS data are also included for comparison. The \xmm\ pn, MOS and \suzaku\ XIS data have been analyzed by \citet{berna2013}. We do not include \swift\ observations in our analysis because the observations with XRT and BAT were not taken at the same time. In Table 1 we list the observations used in this study. 

\begin{table}
  \caption{Observations used in this work.}
  \label{tbl:observations}
 \begin{center} \begin{tabular}{ccc}
    \hline
    Telescope & Obs. ID & Effective Exposure \\
    &  & (ks) \\
    \hline
    \xmm\ & 0671850301 & 13.6/18.9$^{a}$\\
\suzaku\ & 406042010 & 39.5\\
    \hline
  \end{tabular}
  \end{center}
Note: a. The effective exposures of pn/MOS instruments, respectively.\\
\end{table}

We reprocess the \suzaku\ data downloaded from the \suzaku\ archive, using the software package heasoft (version 12.10.0; \citealt{arnaud1996}). The raw data are reduced with the standard routine \textit{aepipeline} in xspec with the latest calibration files (XIS:20181010). We extract the XIS on-source spectrum and an off-source background spectrum, together with the response matrix (rmf) and effective area (arf) files, using \textit{xselect}. The source region is a circular region centered on the source, with a radius R = 220\arcsec, while the background region is an annulus with the inner and outer radii equal to $250\arcsec$ and $400\arcsec$, respectively. The background files of the HXD data are downloaded from the Suzaku background FTP server\footnote{ftp://legacy.gsfc.nasa.gov/suzaku/data/background}, and processed using the \textit{hxdpinxbpi} tool. All spectra are regrouped to ensure a S/N ratio of three per bin.

We also reprocess the \xmm\ data downloaded from the \xmm\ archive, using ``emproc" (MOS1, MOS2) and ``epproc" (PN) commands in the Science Analysis System (SAS, v15.0.0) software. We exclude time intervals of flares when the counts rate above the $10$keV energy band is higher than 0.35 cts/s and 0.8 cts/s for MOS and PN, respectively. Source events are extracted from a $40\arcsec$ circle centered at the source, and the background events are extracted from a $40\arcsec$ circle on the same chip, beside the source region. The response files and ancillary files are generated with the tools \textit{rmfgen} and \textit{arfgen}, respectively. All spectra are again regrouped to ensure a S/N ratio of three per bin.

To constrain \tmax, we fit the 3-10 keV XIS and 12-50 keV HXD background-subtracted spectra with the model $constant\times~pha\times~pcfabs~\times~(mkcflow+Gaussian)$, where \textit{pha} and \textit{pcfabs} respectively represent the foreground and intrinsic absorption, \textit{Gaussian} the Fe I-$K\alpha$ line centered at $6.4$ keV, and \textit{mkcflow} the rest of the spectra. Here $constant$ is a renormalization factor, and is set to $1$ and $1.16$ for XIS and HXD spectra, respectively (see Suzaku Memo 2008-06\footnote{ftp://legacy.gsfc.nasa.gov/suzaku/doc/xrt/ suzakumemo-2008-06.pdf}). The foreground absorption is fixed at $0.103\times10^{22}$cm$^{-2}$, following \citet{berna2013}.

We also jointly fit the 5-8 keV spectra extracted from \suzaku\ XIS0, XIS1, XIS3, \xmm\ pn, MOS1 and MOS2 instruments to constrain the Fe XXVI to Fe XXV line flux ratios (\iratio). The spectra are fitted with the model $pha\times~(brem+threegau)$, where $brem$ stands for the bremsstrahlung model, and the \textit{threegau} were built by \citet{xu2016} to measure \iratio\ values.

\section{Results}
In the upper panel of Figure 1 we present the \suzaku\ XIS and HXD spectra with the best fitted model, and the 5-8 keV spectra from \suzaku\ XIS, \xmm\ pn and MOS instruments are presented in the lower panel. All spectra are well fitted, judged by the $\chi^{2}$ values ($1.11/931$ and $0.96/1297$, respectively). The best fitting parameters are listed in Table 2. The fitting yields an un-absorbed 0.3-50 keV flux of $1.0\times10^{-10}$ erg s$^{-1}$ cm$^{-2}$ (or a luminosity of $5.3\times10^{33}$ erg s$^{-1}$), or a mass accretion rate of $1.2\times10^{-9}M_{\odot}$yr$^{-1}$assuming the X-ray emission efficiency $\eta=0.5$.

\begin{figure}
  \centering
 \includegraphics[scale=0.34,angle=270]{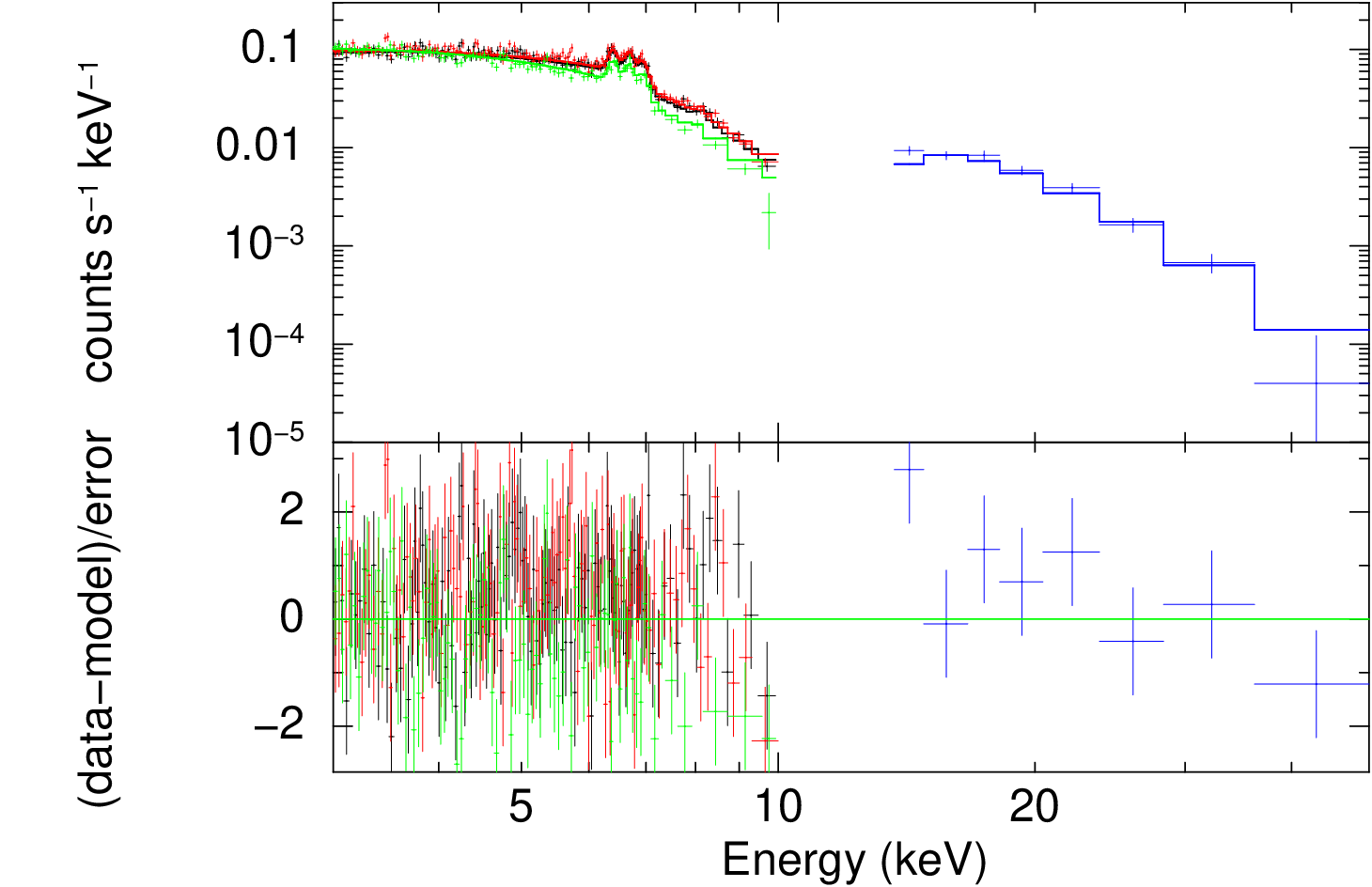}
 \includegraphics[scale=0.34,angle=270]{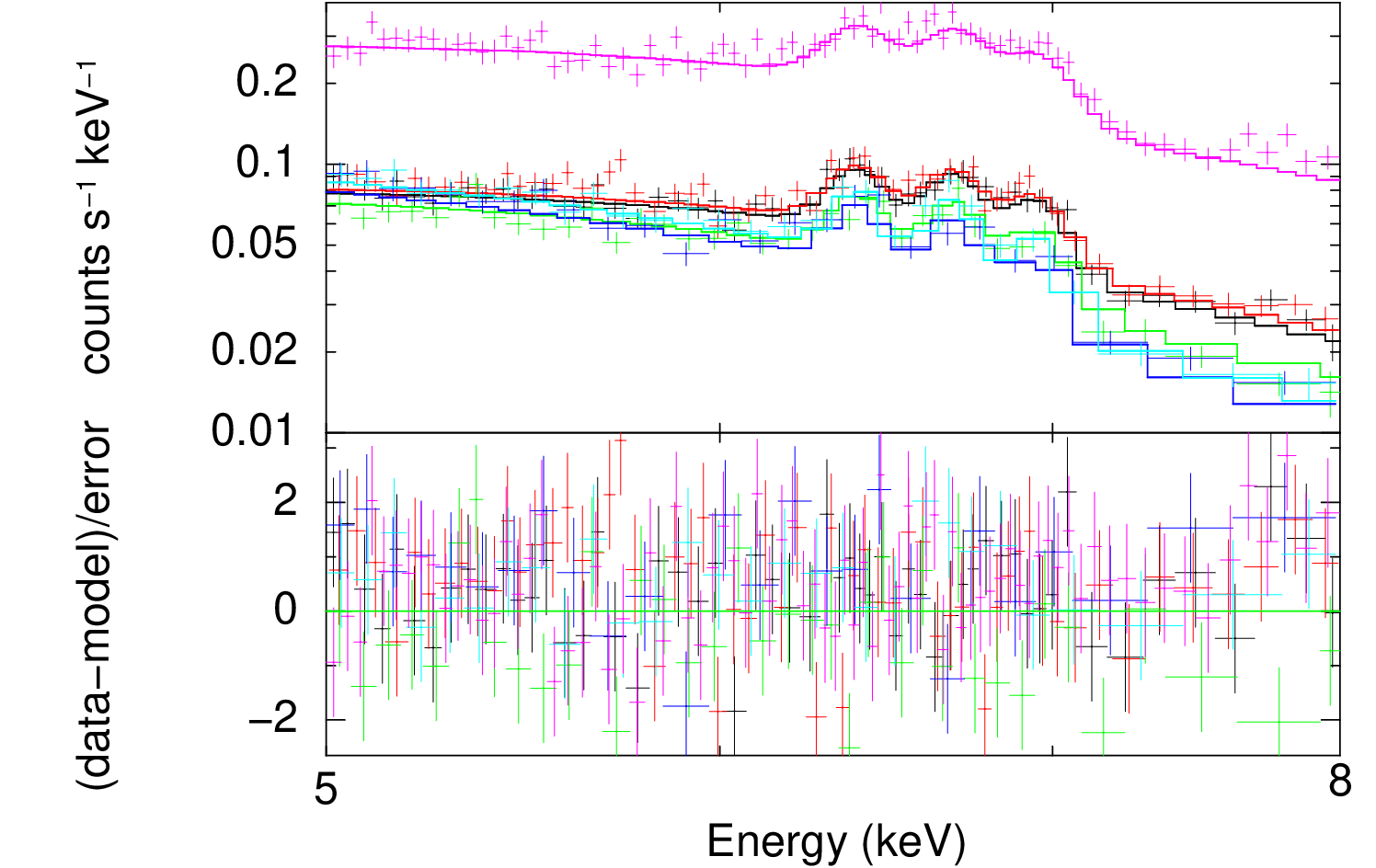}
  \caption{Spectra and best fitting models from \suzaku\ and \xmm\ observations. Upper panel: \suzaku\ XIS (black, green and red for XIS0, XIS1, and XIS3, respectively) and HXD spectra jointly fitted with an absorbed mkcflow model; Lower panel: \suzaku\ XIS (black, green and red for XIS0, XIS1, and XIS3, respectively), \xmm\ pn (purple) and MOS (blue and cyan for MOS1 and MOS2, respectively) spectra jointly fitted with an absorbed bremsstrahlung with threegau model to account for the Fe lines. The spectra have been re-binned for plotting only.}
\label{fig:xray}
\end{figure}

\begin{table}
  \caption{Spectral parameters of the best-fitting models. Uncertainties are given in 90\% level.}
  \label{tbl:fitting}
 \begin{center} \begin{tabular}{ccc}
    \hline
	\multicolumn{3}{c}{\suzaku\ XIS 3--10 keV \& HXD 12--50 keV spectra$^{a}$}\\

    \hline
$N_{\rm H}$ & $10^{22}$cm$^{-2}$ & $0.103$\\
$N_{\rm H,pc}$ & $10^{22}$cm$^{-2}$ & $35\pm3.1$\\
cvf & percent & $0.5\pm0.03$\\
$T_{\rm max}$ & keV & $35_{-9}^{+14}$\\
$\chi_{\nu}^2$(d.o.f) & & $1.11(931)$\\
\hline
	\multicolumn{3}{c}{\suzaku\ XIS, \xmm\ pn \& MOS 5--8 keV spectra$^{b}$ }\\
\hline
$N_{\rm H}$ & $10^{22}$cm$^{-2}$ & $17.4\pm6.8$ \\
$T$ & keV & $23.9_{-6.2}^{+16}$\\
\iratio\ & & $0.74\pm0.09$ \\
$\chi_{\nu}^2$(d.o.f) & & $0.96(1297)$ \\
\hline
  \end{tabular}
  \end{center}
Note: $^{a}$: The \suzaku\ XIS and HXD spectra are fitted with an absorbed mkcflow model.\\
$^{b}$: The \suzaku\ XIS, \xmm\ pn and MOS spectra are fitted with an absorbed bremsstrahlung with threegau model.\\
\end{table}

The WD mass can then be inferred to be $0.77\pm0.11M_{\odot}$ using the measured $T_{\rm max}$ ($35_{-9}^{+14}$ keV) and the $T_{\rm max}$--$M_{\rm WD}$ relation for IPs \citep[e.g.,][]{mukai2017,yu2018,xu2019}. What's more, the \iratio\ value can also be used to put constrains on the $T_{\rm max}$ and $M_{\rm WD}$, as suggested by \citet{xu2016} and \citet{xu2019}. The predicted value of $T_{\rm max}=33\pm6$ keV (or a $M_{\rm WD}=0.75\pm0.07M_{\odot}$) is consistent with that from the broad band spectral fitting, which shows the reliability of our measurements. In the rest of this work we take $0.77\pm0.11M_{\odot}$ as the WD mass.

We estimate the companion mass to be $0.55\pm0.11M_{\odot}$ from the mass ratio $q=M_{\rm WD}/M_{\rm 2}=1.42\pm0.2$ (assuming that the companion fills its RL, \citealt{tov2018b}), with all the uncertainties considered.  The mass measurements, together with previous orbital period and radial velocity results \citep[e.g.,][]{tov2018a,tov2018b}, implies an inclination angle of $i=18^{\circ}$.

\section{Simulation}

To evaluate the reliability of the above solution in detail, we perform numerical simulation of the binary evolution with the Modules for Experiments in Stellar Astrophysics (MESA; version number 11554, \citealt{paxton2011,paxton2013,paxton2015,paxton2017}) code to check if it can reproduce the observational characteristics of V1082. In the simulation we adopt Solar metallicity ($Z=0.02$). 

We assume that magnetic braking works in stars less massive than 1.4 $M_{\odot}$ which have a convective envelope \citep {verbunt1981,rappaport1983,lin2011,kalo2016}. All other parameters are set to take their default values in the code. Since the retention efficiencies for both hydrogen and helium burning are still highly uncertain, for simplicity, we assume the WD does not accumulate the accreted matter, and the lost matter carries the specific angular momentum of the WD \citep{kalo2016}. We set the initial WD mass $M_{\rm WD}$ to be  $0.77M_{\odot}$ and $0.88M_{\odot}$, the initial companion mass $0.9M_{\odot}\leq M_{2} \leq 2.6M_{\odot}$ (with a step of $0.1M_{\odot}$), and the initial orbital period  $0.5\leq P_{\rm orb} \leq 5$ days (with a step of $0.05$ day). To compare with previous works, we also set $M_{\rm WD}$ to be $0.64M_{\odot}$ \citep{berna2013} and perform simulation accordingly.

\begin{figure}
  \centering
 \includegraphics[scale=0.35]{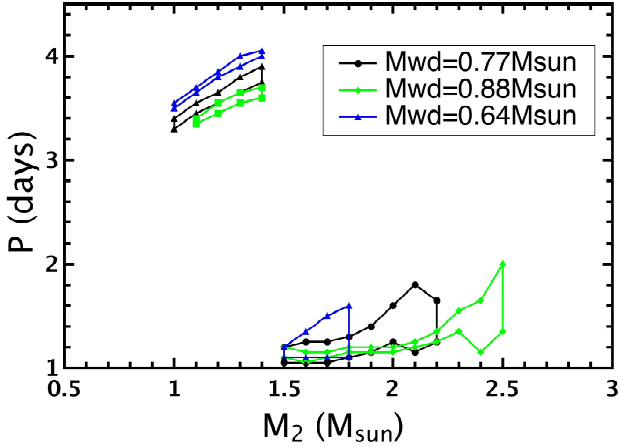}
  \caption{The initial $P_{\rm orb}$--$M_2$ parameter spaces for possible progenitors of V1082. Systems with initial $P_{\rm orb}$ and $M_2$ inside the polygons could evolve to CVs with $P_{\rm orb}=0.86$ day and $M_2=0.55\pm0.11M_{\odot}$. The black, green and blue polygons refer to $M_{\rm WD}=0.77M_{\odot}$, $0.88M_{\odot}$ and $0.64M_{\odot}$, respectively. Magnetic braking are initially turned on for binaries with $M_{2}\leq1.4M_{\odot}$ (binaries inside the upper-left polygons), and off for others (binaries inside the lower-right polygons).}
\label{fig:initial}
\end{figure}

\begin{figure}
  \centering
 \includegraphics[scale=0.3]{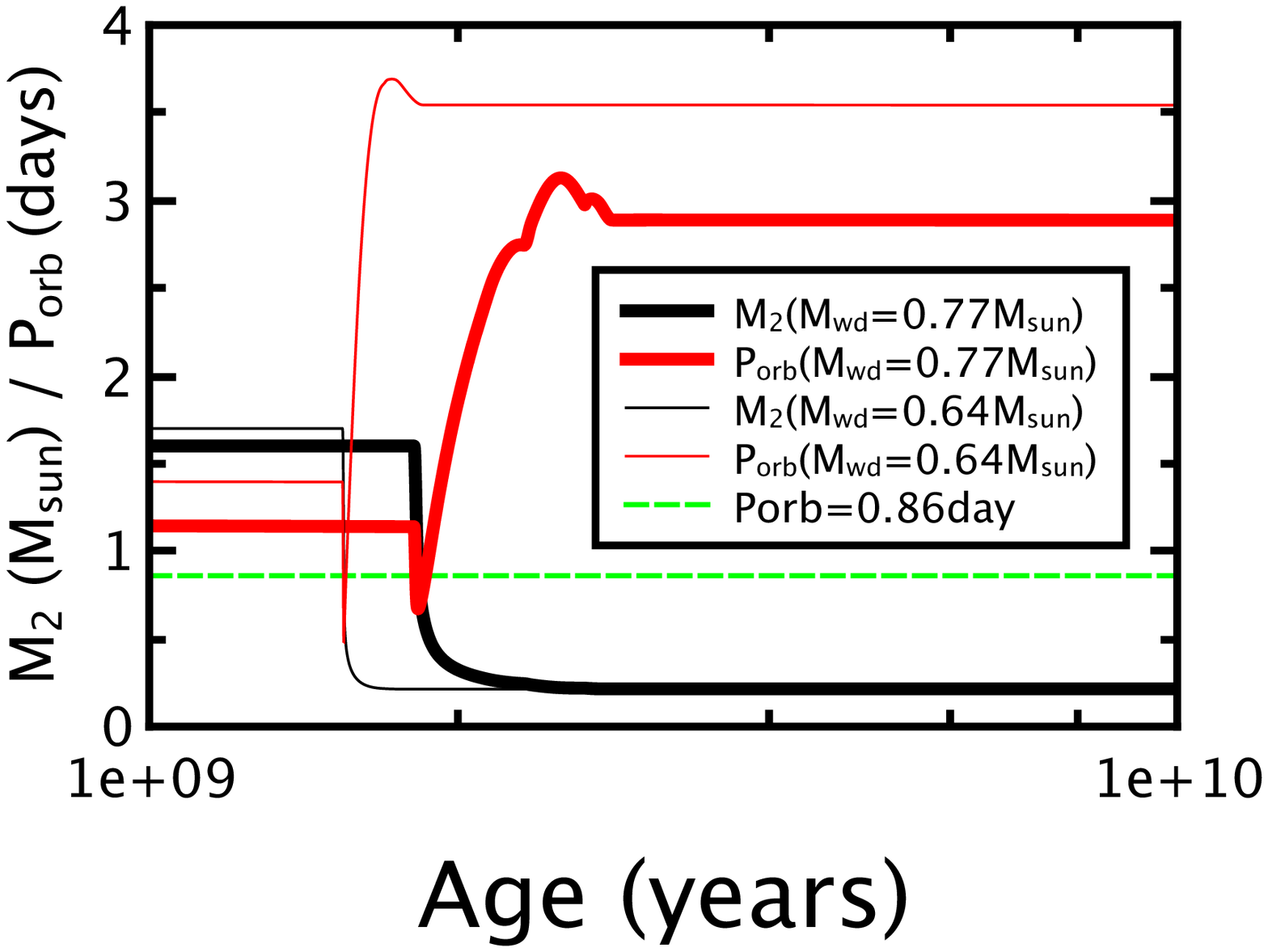}
 \includegraphics[scale=0.3]{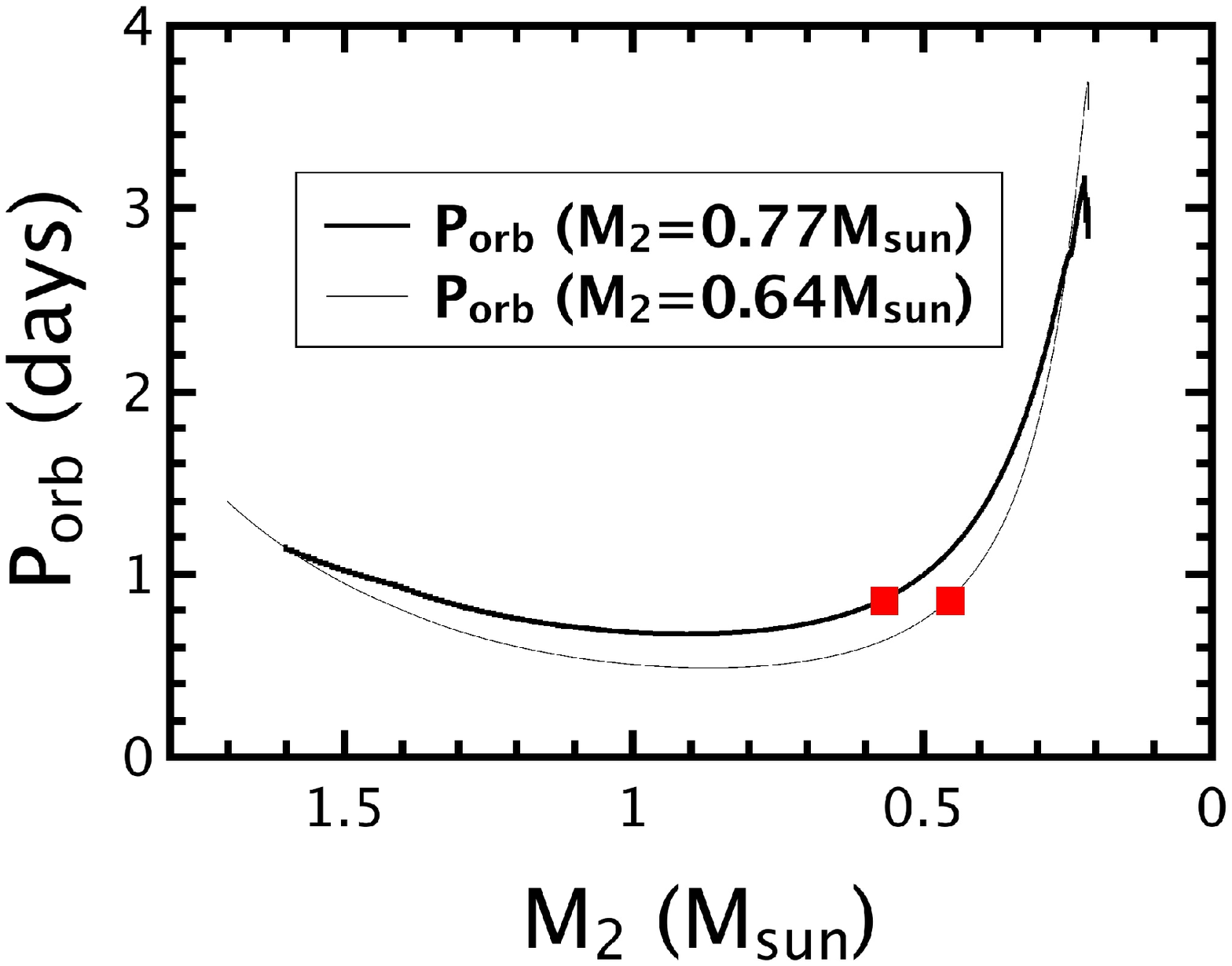}
\includegraphics[scale=0.3]{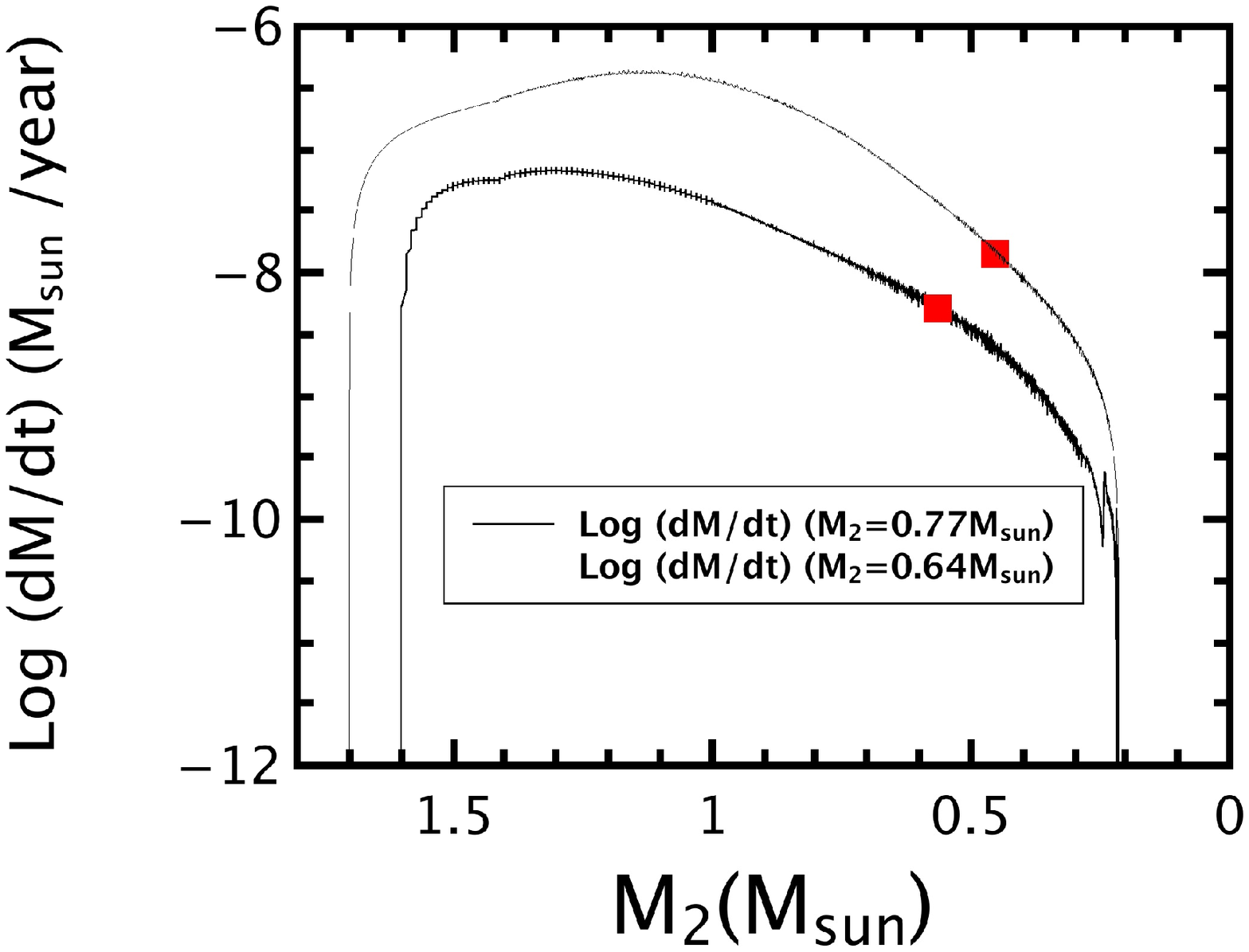}
\includegraphics[scale=0.3]{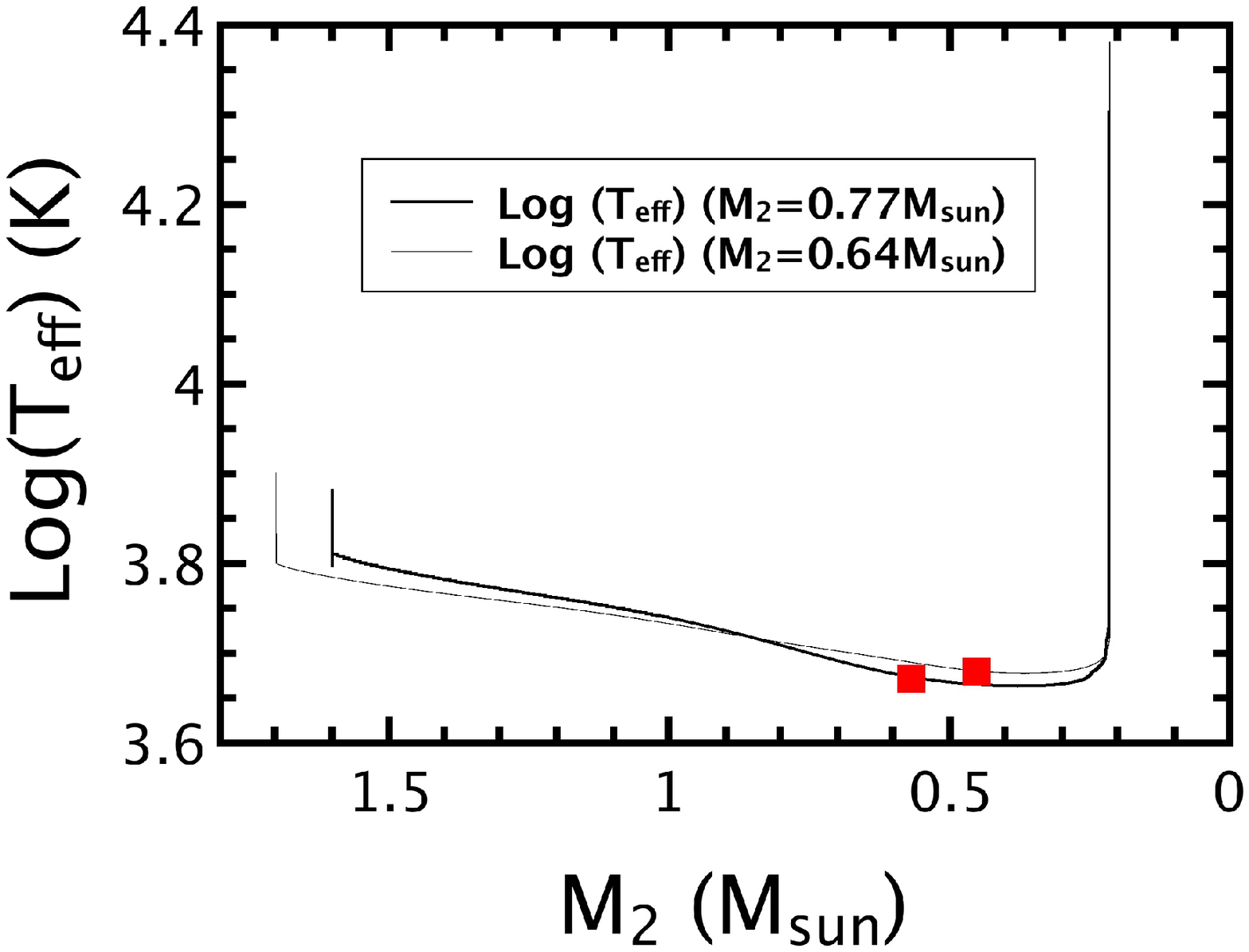}
  \caption{Two possible evolutionary paths of V1082. The first one is shown with thick solid lines. The system initially contains a $0.77M_{\odot}$ WD and a $1.6M_{\odot}$ zero-age MS companion in a $1.15$ day circular orbit. The first panel shows the orbital period and the mass of the donor star as a function of stellar age. The rest three panels show the binary orbital period, the mass transfer rate and the effective temperature of the donor star as a function of the donor mass. In all the panels, the system begins its evolution from left and evolves to the right. The second one is shown with thin solid lines. It starts with a $0.64M_{\odot}$ WD and a $1.7M_{\odot}$ zero-age MS companion in a $1.4$ day circular orbit. Its evolution is similar to the first one. The red squares represent when the binary evolves to a V1082--like CV (with $P_{\rm orb}=0.86$ day, $M_{2}=0.568M_{\odot}$, and $T_{\rm eff}=4714$ K for the first path, and $M_{2}=0.45M_{\odot}$ and $T_{\rm eff}=4787$ K for the second path), as denoted in the figures. The apparent `perturbation' on the $dM/dt$--$M_2$ diagram are only numerical effects. }
\label{fig:evolution}
\end{figure}

In Figure 2 we present the distribution of the possible progenitors of V1082 in the $P_{\rm orb}$--$M_2$ plane, with the black, green and blue polygons for $M_{\rm WD}=0.77M_{\odot}$, $0.88M_{\odot}$ and $0.64M_{\odot}$, respectively. As shown in the figure, there are two separated allowed parameter spaces for each WD mass. The one on the upper left represents binaries with initial  $M_{2}\leq1.4M_{\odot}$, where magnetic braking is turned on at the onset of their evolution. These systems have initial $M_2$ of 1.0--1.4 $M_{\odot}$ and $P_{\rm orb}$ of 3.2--4.1 days. The other one on the lower right represents binaries with initial $M_{2}>1.4M_{\odot}$, where the magnetic braking is turned on when the companion has lost some mass so that $M_{2}<1.4M_{\odot}$. Such systems have initial $M_2$ of 1.5--2.5 $M_{\odot}$ and $P_{\rm orb}$ around 1--2 days. Binaries in these regions have a companion with $T_{\rm eff}\sim4660$--$5100$ K when they evolve to the V1082--like (i.e., with $P_{\rm orb}=0.86$ and $M_2\sim 0.55\pm0.11M_{\odot}$) systems. Their mass transfer rates are  $\sim 10^{-9}$--$10^{-7}M_{\odot}$yr$^{-1}$, enough to power the observed X-ray luminosity. Binaries outside the polygons could not evolve to V1082--like CVs. They either experience unstable mass transfer or evolve to systems with parameters inconsistent with those of V1082.

Two example evolutionary paths of V1082 are presented in Figure 3. The first one is shown with solid lines, and the second one is shown with dashed lines. The first simulation starts with a $0.77M_{\odot}$ WD and a $1.6M_{\odot}$ zero-age MS companion in a $1.15$ day circular orbit. As shown in Figure 3, the companion evolves off MS and fills its RL at the age of $1.8\times10^9$ years. Because the companion is more massive than the WD, mass transfer proceeds on a thermal timescale at a rate around several $10^{-8}M_{\odot}$yr$^{-1}$, and the binary orbit shrinks accordingly. After 20 Myrs of mass transfer, the companion has lost $\sim0.7M_{\odot}$ mass, and the orbit period reaches its minimum of $0.67$ day. The mass transfer rate continues to drop after that, and the orbit starts to enlarge. At $1.86\times10^9$ yrs, the companion becomes a $0.568M_{\odot}$ star in a $0.86$ day orbit, with effective temperature of $4714$ K, radius of $1.48R_{\odot}$ and luminosity of $0.97L_{\odot}$. Its V band magnitude can be estimated to be 14.8, considering the interstellar reddening of $E(B-V)=0.15$ \citep{schlegel1998}. All these values are consistent with the observational ones \citep[e.g.,][]{tov2018b}. The mass loss rate of the companion at present is $5.3\times10^{-9}M_{\odot}$yr$^{-1}$, well above the requested value to power the X-ray luminosity of V1082. The second simulation starts with a $0.64M_{\odot}$ WD and a $1.7M_{\odot}$ zero-age MS companion in a $1.4$ day circular orbit, and its evolution path is similar to that of the first one. At $1.554\times10^9$ years, the companion becomes a $0.45M_{\odot}$ star in a 0.86 day orbit, with an effective temperature of $4787$ K, radius of $1.37R_{\odot}$, luminosity of $0.90L_{\odot}$ and a companion mass loss rate of $1.5\times10^{-8}M_{\odot}$yr$^{-1}$. These values are also consistent with the observational results. We then conclude that a mass transferring mCV can well reproduce most of the observational facts of V1082.

\section{Discussion \& Summary}
We propose an IP model with a more massive WD and a RL--filling companion to explain the high luminosity of V1082. Our numerical simulation shows that the companion has evolved off MS and lost $\sim0.5$--$2.0M_{\odot}$ mass in its past evolution. With an effective temperature of $4660$--$5100$ K, the companion appears as a K type star, and is losing mass at a rate of $\sim10^{-9}$--$10^{-7}M_{\odot}$yr$^{-1}$. Such a scenario can well account for the orbital period, the high X-ray luminosity and the apparently K type companion. In the following we compare our results to previous works and other observations. 

Firstly, our proposed WD mass is a little bit higher, although still compatible with the one determined by  \citet{berna2013}, $0.64\pm0.04M_{\odot}$. We speculate that this discrepancy could originate from the non-simultaneous observations with \xmm\ and \swift\ BAT in their work. As pointed by \citet{berna2013}, V1082 showed strong variability by a factor up to $\sim3-100$ between \xmm\ and \swift\ XRT observations. Such a variation may considerably influence the derived value of $M_{\rm WD}$. However, our model does not strongly depend on the exact WD mass values. As shown in Section 4, binaries with either a $0.77M_{\odot}$ or a $0.64M_{\odot}$ WD can well explain the observations, as long as it has a RL filling, off-MS companion. Our simulations also suggest that the initial donor masses of V1082 have to be set to different values if different WD masses (e.g., $0.77M_{\odot}$ and $0.64M_{\odot}$, see Figure 3 for details) are adopted. In such way, the progenitor binary could evolve to a RL--filling IP whose mass ratio is consistent with $q=1.42\pm0.2$, as suggested by previous observations \citep{tov2018b}. On the other hand, if we adopted the donor mass of $0.73\pm0.04M_{\odot}$ \citep{tov2016, tov2018b}, the simulation shows that the donor star would still be in main sequence after $1.4\times10^{10}$ years. The radius of the star is $R=0.72R_{\odot}$, which is significantly less than its RL radius ($R_{\rm RL}\sim1.6R_{\odot}$, assuming a $0.6$--$0.8M_{\odot}$ compact companion star), and could not enable a mass transfer rate of $\sim10^{-9}M_{\odot}$yr$^{-1}$ as requested by the X-ray observations \citep{berna2013}.

Secondly, our proposal assumes a RL--filling companion, which is in contrast to the RL under-filling companion, pre-polar model by \citet{tov2018b}. \citet{tov2018a} reported that the light curve (LC) of V1082 in its deep minimum did not show the double-hump structure, and interpreted that the companion star is not ellipsoidal shaped, so is not filling its RL. However, the non-detection of double-hump structure in the LC could also be the result of the low inclination angle. We make numerical simulations with the ELC code \citep{orosz2000} to examine this possibility. Setting $i=18^{\circ}$, the results show that there is indeed a new peak in the LC with an amplitude of $\sim0.012$ magnitude. We check the LC shown in Figure 9 from \citet{tov2018a}, and find that the amplitude of the LC variation is $\sim0.12$ magnitude, which means the new peak would cause a $\sim10\%$ variation in the LC. The reason why such a new peak was not identified by \citet{tov2018a} is not clear yet, and we tentatively suggest two possible explanations based on the LC used in \citet{tov2018a} as follows.
Firstly, to reduce contaminations from the accreting WD, the LC analysis was based on the K2 data when the system reached deep minimum. However, the quality of the LC data was still not good enough to allow a serious modeling, as described in \citet{tov2018a}. The available data only cover two and a half orbital cycles so  a quantitative fitting was not possible \citep{tov2018a}. As a result, it is difficult to identify a new peak. Secondly, the profiles of the variation are changing in the two and a half cycles, and the fluctuations in the LC already exceed $0.12$ magnitude. For example, there are 0.02 magnitude differences in the peak magnitudes in the orbital phase of $0.4$ and $1.4$), and the fluctuations of the observed data points reach 0.2 magnitude in the orbital phase $-0.1$ and $0.7$. The possible second peak might have been missed for such large fluctuations. We strongly suggest further observations to better constrain the LC of the companion star and the nature of V1082. What's more, the low inclination angle means that one of the magnetic poles of the white dwarf might always be visible, thus may explain why there is no detection of the spin of the white dwarf. \citet{tov2018b} also suggested that the position of the high amplitude He II 4686 line component is somewhere near the WD in the Doppler map. In our model, this component may arise from the  inner most part of the accretion disk, where the accreted matter is illuminated by the X-rays from the WD. 

Thirdly, we noticed that the system showed strong variability (up to two orders of magnitude) in past observations \citep{berna2013}. What's more, the white dwarf's accretion rate ($1.2\times10^{-9}M_{\odot}$yr$^{-1}$) derived from \textit{Suzaku} observations is lower than the companion's mass loss rate predicted by the stellar evolution code for most systems ($\sim10^{-9}$--$10^{-7}M_{\odot}$yr$^{-1}$). Such phenomena seem usual in IPs with orbital period above 6 hours, as reported by \citet{suleimanov2019}.
The strong variability and the lower accretion rate of the system could be caused by the propeller mechanism. Fig. 2 shows that, at the beginning stage of mass transfer, the WD accretes rapidly, so its spin is also accelerated. After that the mass transfer rate declines, and the magnetospheric radius can be larger than the so-called co-rotation radius, and the WD enters the propeller stage \citep{frank2002}. Most of the inflowing matter is halted at the magnetospheric radius by centrifugal forces, resulting in a significant reduction of the accretion luminosity of the WD. The `on' and `off' states of the accretion refer to the high and low optical phases of the system, respectively. This feature seems not unusual for IPs \citep{campana2018}. 

\section*{Acknowledgments}

The authors thank the anonymous referee for constructive comments that helped improve this paper. This work is supported by the Natural Science Foundation of China under grant Nos. 11873029, 11333004, and 11773015, Project U1838201 supported by NSFC and CAS, and the National Key Research and Development Program of China (2016YFA0400803).

\clearpage

\end{document}